\magnification\magstep1
\baselineskip=12pt
\parindent=3truepc
\def\bigskip{\vskip 6 truemm}
\def\medskip{\vskip 4truemm}

\input epsf.tex
\font\rfont=cmr10 at 10 true pt
\def\ref#1{$^{\hbox{\rfont {#1}}}$}


\font\tenit=cmti10 scaled\magstep0

\def\a {\alpha}  \def\g {\gamma} 
\def\e{\epsilon}  \def\o{\omega}
\def\s {\sigma}

\def\pmb#1{\setbox0=\hbox{#1}
 \kern.05em\copy0\kern-\wd0 \kern-.025em\raise.0433em\box0 }

\def \half {{\scriptstyle {1 \over 2}}}

 %


\def\boxit#1{\vbox{\hrule\hbox{\vrule\kern1pt\vbox
{\kern1pt#1\kern1pt}\kern1pt\vrule}\hrule}}

\def\h{\hfill\break}
\parskip=6pt
\parindent=0pt
\hsize=6truein\hoffset=-5truemm
\voffset=-1truecm\vsize=8.5truein
\def\footnoterule{\kern-3pt
\hrule width 17truecm \kern 2.6pt}
\def\h{\hfill\break}


\catcode`\@=11 

\def\nolabels{\def\wrlabeL##1{}\def\eqlabeL##1{}\def\reflabeL##1{}}
\def\writelabels{\def\wrlabeL##1{\leavevmode\vadjust{\rlap{\smash%
{\line{{\escapechar=` \hfill\rlap{\sevenrm\hskip.03in\string##1}}}}}}}%
\def\eqlabeL##1{{\escapechar-1\rlap{\sevenrm\hskip.05in\string##1}}}%
\def\reflabeL##1{\noexpand\llap{\noexpand\sevenrm\string\string\string##1}}}
\nolabels
\global\newcount\refno \global\refno=1
\newwrite\rfile
\def\defref{$^{{\hbox{\rfont \the\refno}}}$\nref}
\def\nref#1{\xdef#1{\the\refno}\writedef{#1\leftbracket#1}%
\ifnum\refno=1\immediate\openout\rfile=refs.tmp\fi
\global\advance\refno by1\chardef\wfile=\rfile\immediate
\write\rfile{\noexpand\item{#1\ }\reflabeL{#1\hskip.31in}\pctsign}\findarg}
\def\findarg#1#{\begingroup\obeylines\newlinechar=`\^^M\pass@rg}
{\obeylines\gdef\pass@rg#1{\writ@line\relax #1^^M\hbox{}^^M}%
\gdef\writ@line#1^^M{\expandafter\toks0\expandafter{\striprel@x #1}%
\edef\next{\the\toks0}\ifx\next\em@rk\let\next=\endgroup\else\ifx\next\empty%
\else\immediate\write\wfile{\the\toks0}\fi\let\next=\writ@line\fi\next\relax}}
\def\striprel@x#1{} \def\em@rk{\hbox{}} 
\def\lref{\begingroup\obeylines\lr@f}
\def\lr@f#1#2{\gdef#1{\defref#1{#2}}\endgroup\unskip}
\newwrite\lfile
{\escapechar-1\xdef\pctsign{\string\%}\xdef\leftbracket{\string\{}
\xdef\rightbracket{\string\}}}

\def\writestop{\def\writestoppt{\immediate\write\lfile{\string\p
ageno%
\the\pageno\string\startrefs\leftbracket\the\refno\rightbracket%
\string\def\string\secsym\leftbracket\secsym\rightbracket%
\string\secno\the\secno\string\meqno\the\meqno}\immediate\closeout\lfile}}
\def\writestoppt{}\def\writedef#1{}
\catcode`\@=12 

\rightline{DAMTP 96/48}
\bigskip
\centerline{THE SOFT POMERON}
\vskip 10truemm
\centerline{P V LANDSHOFF}
\centerline{\tenit DAMTP, University of Cambridge}
\centerline{\tenit Cambridge CB3 9EW, England}
\vskip 10truemm
\centerline{ABSTRACT}
\vskip -2mm

\midinsert\leftskip 3truepc\rightskip 3truepc{{\rfont\noindent
The soft pomeron successfully correlates a wide variety of data.
Its properties seem rather simple: it couples to single quarks and
its coupling factorises.
}}\endinsert
\bigskip\rm
{\bf 1 Introduction}

The history of the soft pomeron goes back more than 35 years. In the 1960's
a well-defined mathematical theory was developed, based on the idea of
making angular momentum a complex variable, and  there was a 
great deal of successful
but very dirty phenomenology, but there was little or no understanding
of what pomeron exchange is in physical terms.

In the 1970's there was rather little work on the subject; attention turned
instead to hard processes.

In the 1980's data from higher energies
revealed that actually the phenomenology is surprisingly clean.
There were the beginnings of a crude physical understanding, based
on nonperturbative gluon exchange, and there were several successful
predictions.

Now, in the 1990's, HERA is providing important new data and reviving the
interest in the soft pomeron. The hope is that this will lead to a fuller
understanding, but it will surely be the 2000's before we have a good
physical and theoretical understanding of what pomeron exchange actually 
{\it is}.

In studying the pomeron, it is particularly important to remember that
high energy physics is {\it one} subject: we are much more likely to get
an understanding if we correlate information from many reactions --
$ep$, $\bar pp,\dots$. Indeed, we cannot claim any success until we have done
so. Mere parametrisation of data is of little use: we want a  dynamical
understanding. A superb fit with 20 parameters is much less use than a 
reasonable one with only 3, if we want to extract the physics message from
the data. 

For these reasons, my philosophy is to explore how well one can do with the 
simplest assumptions. It is important, though, that they should be
assumptions that do not conflict with known basic principles.
\topinsert
\centerline{{\epsfxsize=15truemm\epsfbox{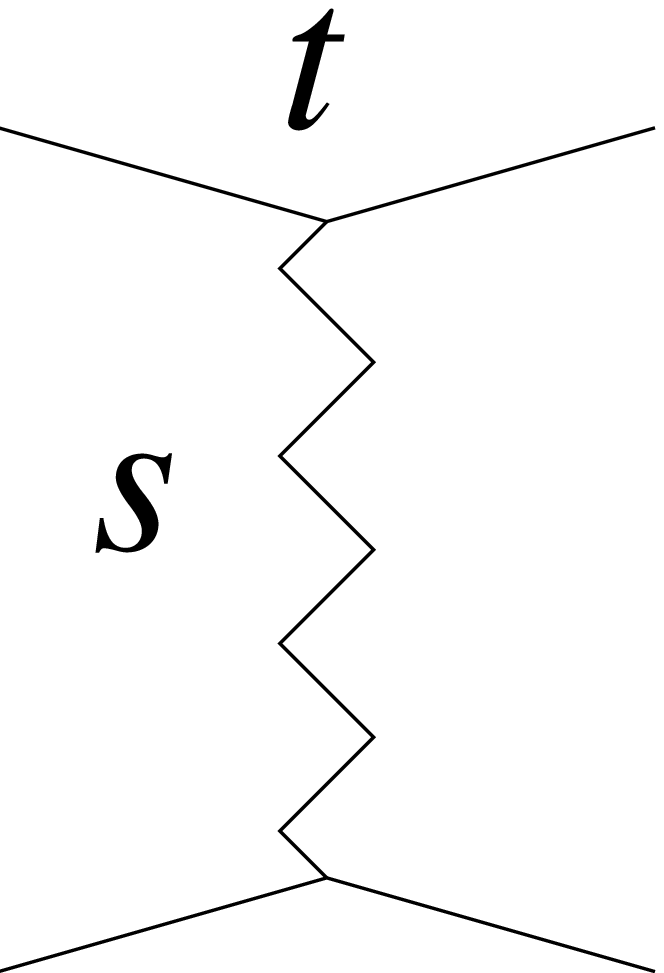}}}\hfill
\vskip -5mm
\centerline{\rfont Figure 1: exchange of a family of particles}
\endinsert

\bigskip
{\bf 2. Complex angular momentum}

A well-defined mathematical formlism, called Regge theory\defref\collins{
P D B Collins, {\sl Introduction to Regge Theory}, Cambridge
University Press (1977)
}, was developed more than 35 years ago to describe the exchanges of families
of particles, for example the spin-one $\rho$ together with its spin-3, spin-5,
$\dots$ excitations. Suppose that these exchanges are in the $t$ channel: see
figure 1. Consider the crossed channel, in which $\surd t$ is the centre-of-%
mass energy, and $\ell$ is the orbital angular momentum. The partial-wave
amplitudes $A(\ell ,t)$ are then defined for $\ell=0,1,2,\dots$. Continue
them to complex values of $\ell$ and introduce the ``$\rho$ trajectory''
$\alpha (t)$ defined  such that
$$
\alpha (m_{\rho}^2)=1,~~~~~~~\alpha (m_{\rho _3}^2)=3,~~~~~~~
\alpha (m_{\rho _5^2})=5,~~~~~~~\dots
\eqno(1)
$$
Experiment finds that $\alpha (t)$ is linear in $t$ and, within the errors,
there are three other families whose trajectories all coincide with that of
the $\rho$. These are the families $\omega, f$ and $a$: see figure 2.
The significance of a trajectory $\alpha (t)$ for a family of particles is
that $A(\ell ,t)$ has a pole in the complex $\ell$-plane:
$$
A(\ell ,t)\sim {1\over\ell -\alpha (t)}
\eqno(2a)
$$
and this gives the amplitude of figure 1 a very simple high-energy behaviour
in the channel where now $\surd s$ is the centre-of-mass energy:
$$
T(s,t)\sim \beta (t) s^{\alpha (t)} \xi_{\alpha (t)}
\eqno(2b)
$$
That is, the amplitude varies with $s$ as a simple power, and has a well-%
defined phase $\xi_{\alpha (t)}$ that varies with the power. The function
$\beta (t)$ is not determined (it comes from whatever multiplies the
pole (2b) in $A(\ell ,t)$, but it is known to be real.

\midinsert
\centerline{{\epsfxsize=90truemm\epsfbox{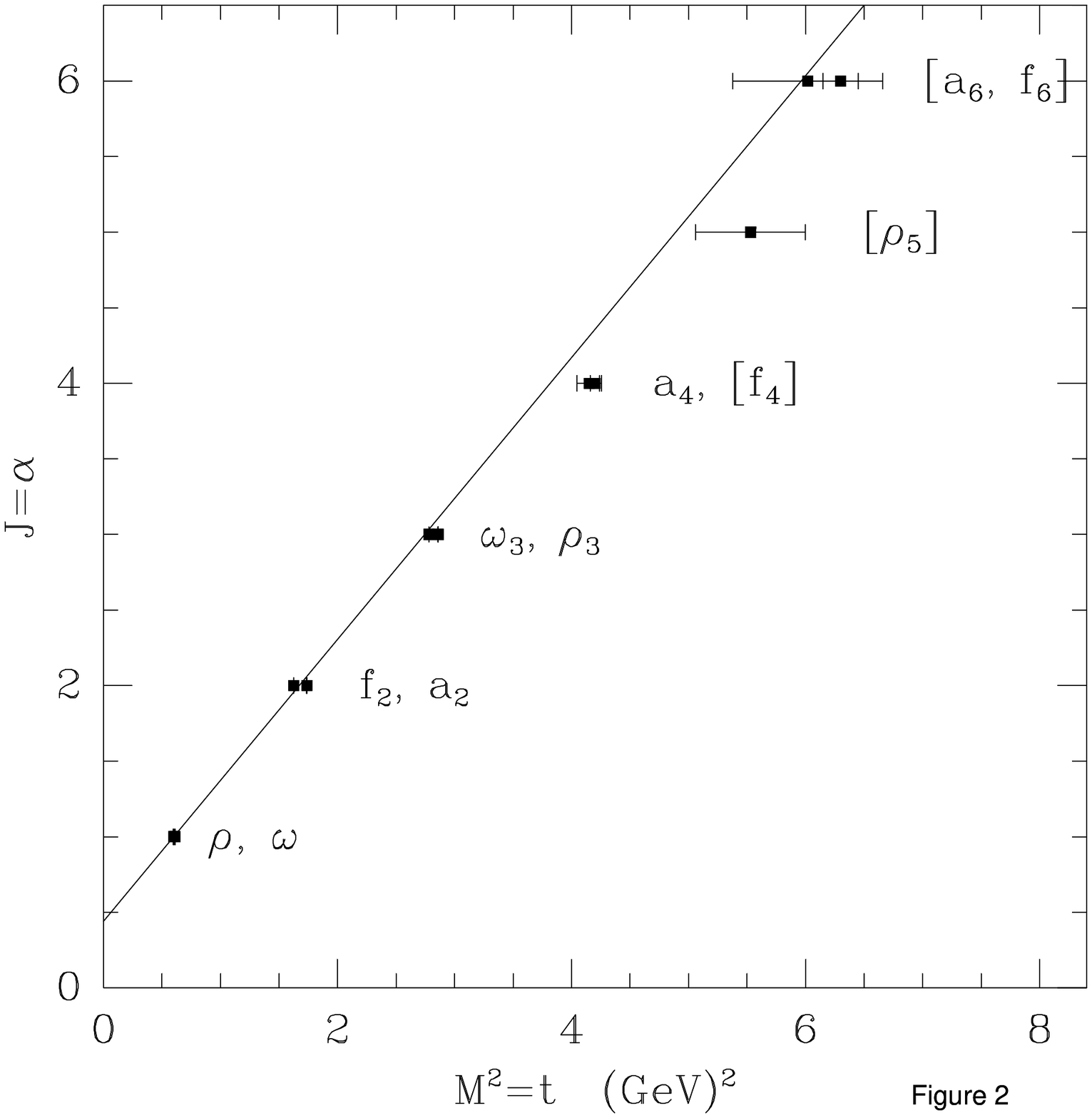}}}\hfill
\vskip -5mm
\centerline{\rfont Figure 2: the $\rho,\omega,f$ and $a$ trajectories}
\endinsert

\topinsert
\vskip -5mm
\centerline{{\epsfxsize=80truemm\epsfbox{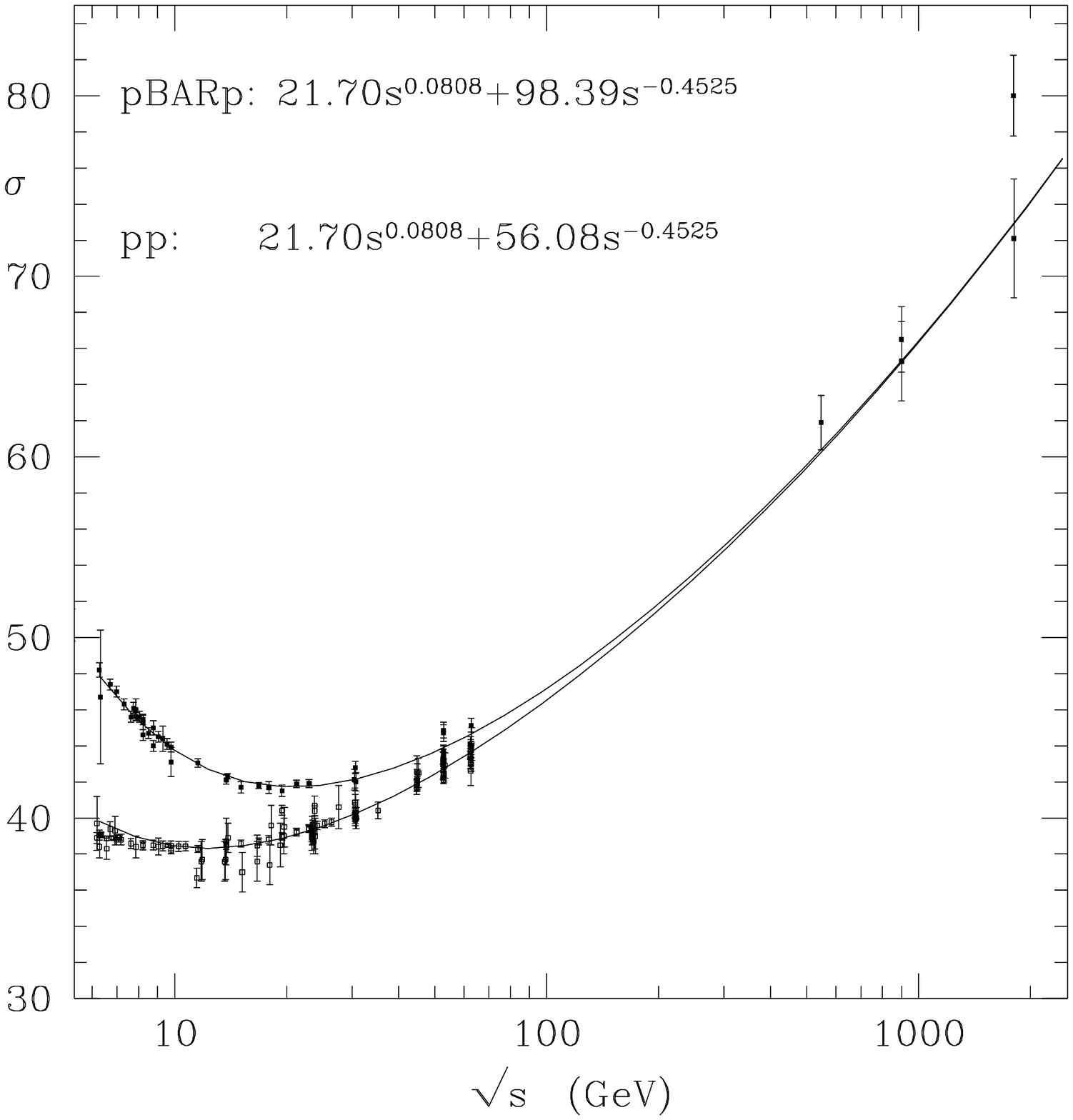}}{\epsfxsize=80truemm\epsfbox{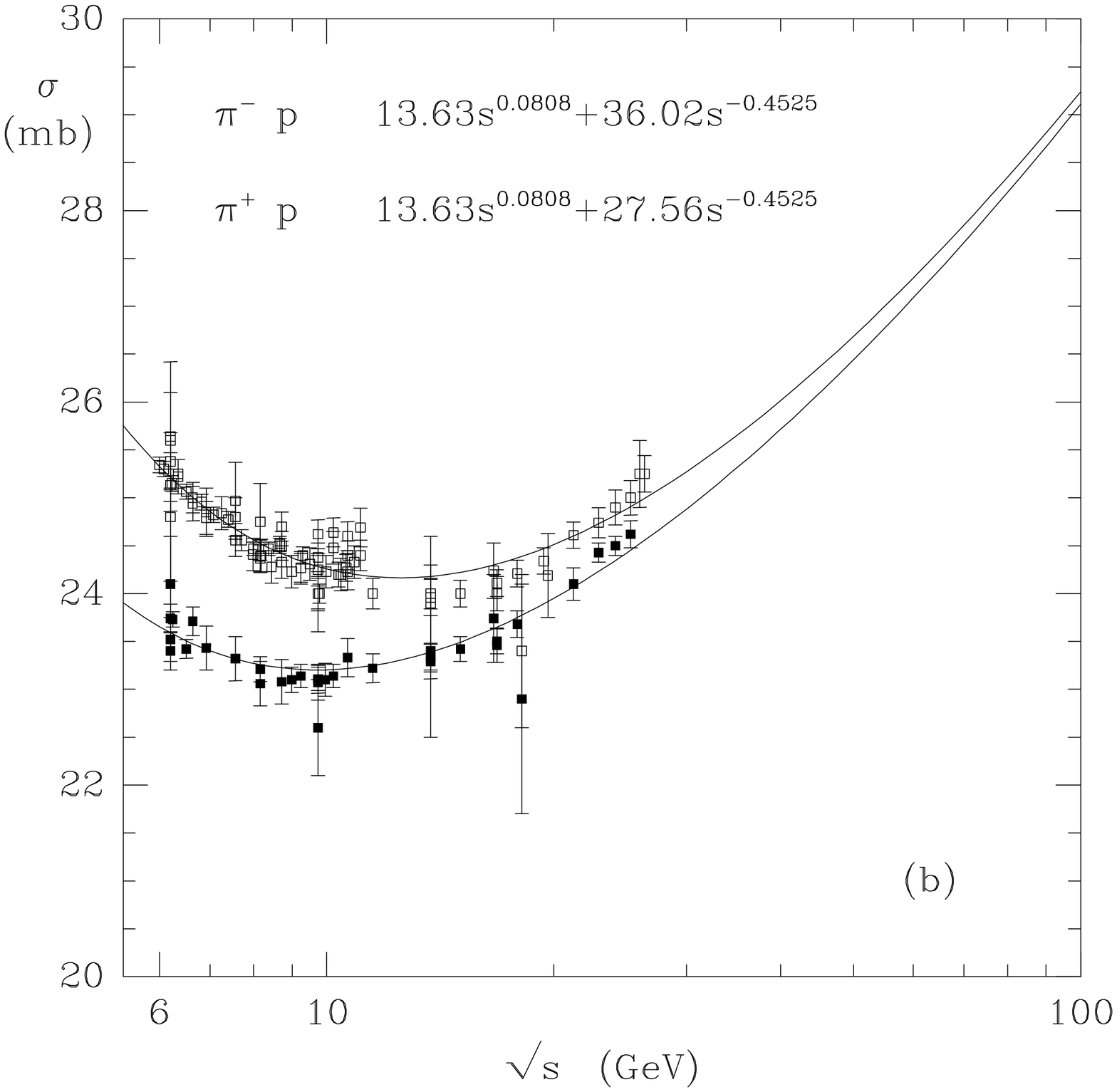}}}\hfill\break
\centerline{{\epsfxsize=80truemm\epsfbox{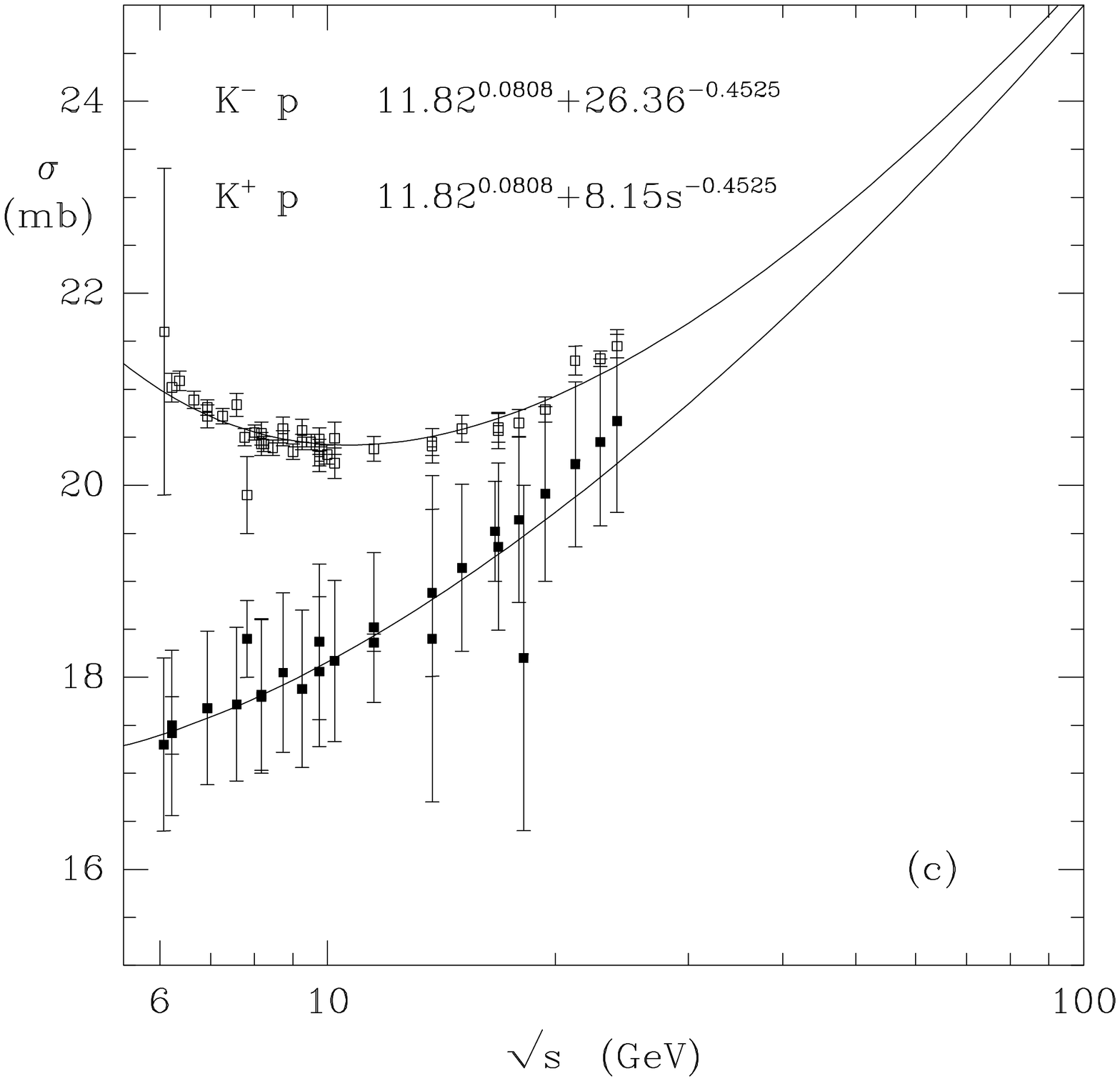}}{\epsfxsize=80truemm\epsfbox{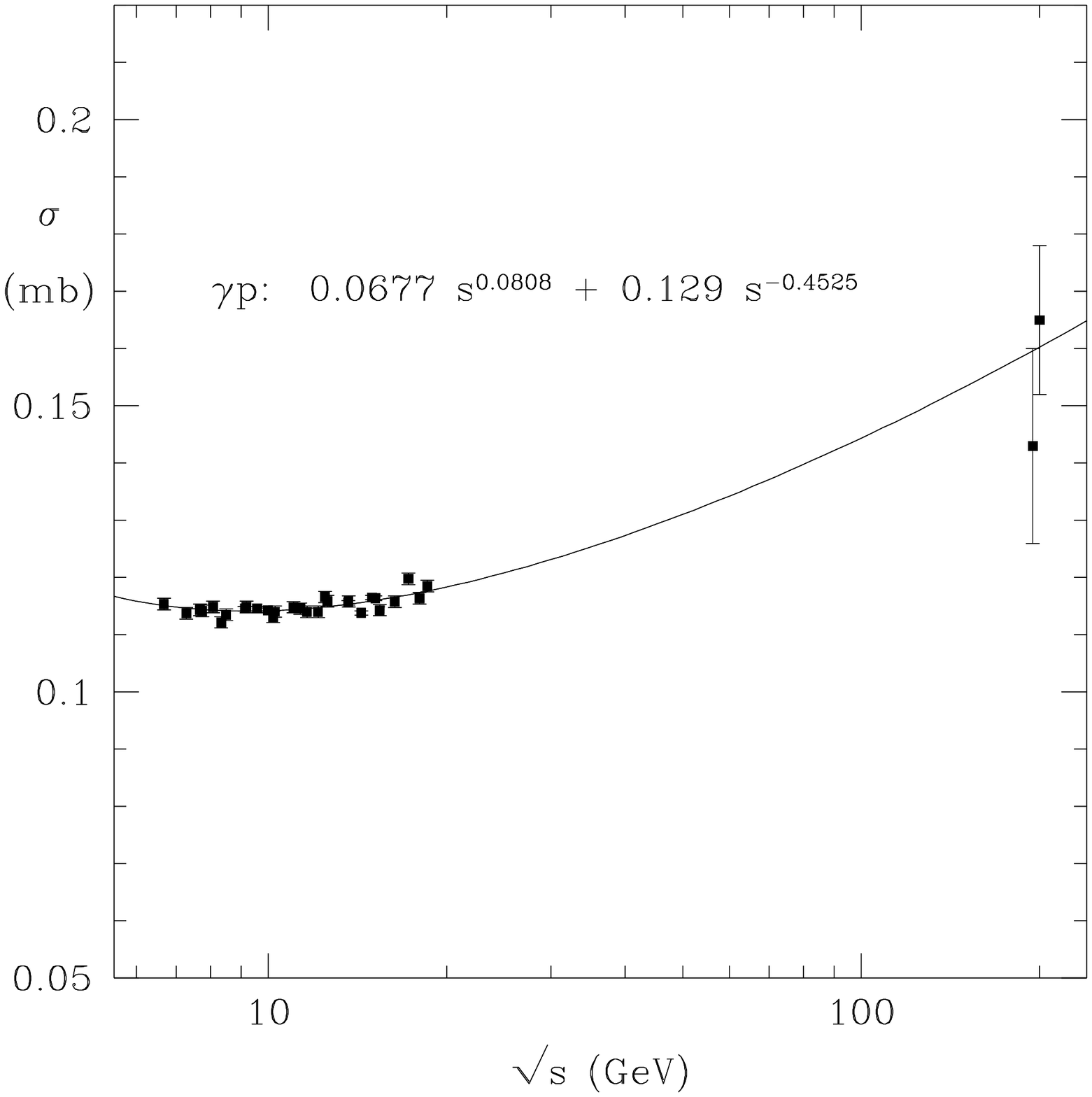}}}\hfill\break
\centerline{\rfont Figure 3: total cross-sections, with simple-power fits
from reference\defref\sigtot{
A Donnachie and P V Landshoff, Physics Letters B296 (1992) 227
}\sigtot}
\endinsert

Unfortunately, it is known that $A(\ell ,t)$ does not only have poles in
the complex $\ell$-plane: there are also branch points. A branch point at
$\ell=\alpha _C(t)$ contributes to the high-energy behaviour of $T(s,t)$
the power $s^{\alpha _C(t)}$, divided by some function of $\log s$ that
depends on just what is the nature of the branch point.

So, to make a high-energy expansion of $T(s,t)$, look for the singularity
in the complex $\ell$-plane with the largest Re $\ell$. This gives the leading
power, together possibly with some log factor. The singularity with the next 
largest Re $\ell$ gives the first nonleading power, and so on. For practical
purposes, that is all: any other ``background'' should be negligible. 
\bigskip

{\bf 3. Total cross-sections and elastic scattering}

From the optical theorem, the total cross-section is
given by
$$\eqalign{
\sigma ^{\hbox{\sevenrm TOT}}&={1\over s}\; \hbox {Im }\;T(s,t=0)\cr
&\sim s^{\alpha (0)-1}\cr}
\eqno(3)
$$
According to figure 2, for $\rho,\omega,f,a$ exchange $\alpha (0) \approx\half$,
so these exchange contribute approximately the power 1/$\surd s$. In order
to describe data, we need also a term that rises slowly with $s$:
see figure 3. The
{\sl simplest assumption} is that this corresponds also to a pole in the
complex $\ell$-plane, and so also gives a simple power of $s$. In order
to give a slowly-rising contribution to $\sigma ^{\hbox{\sevenrm TOT}}$, it should be such
that $\alpha (0)=1+\e _0$ with $\e _0$ a small positive number. We call this
exchange pomeron exchange.

\midinsert
\centerline{{\epsfxsize=50truemm\epsfbox{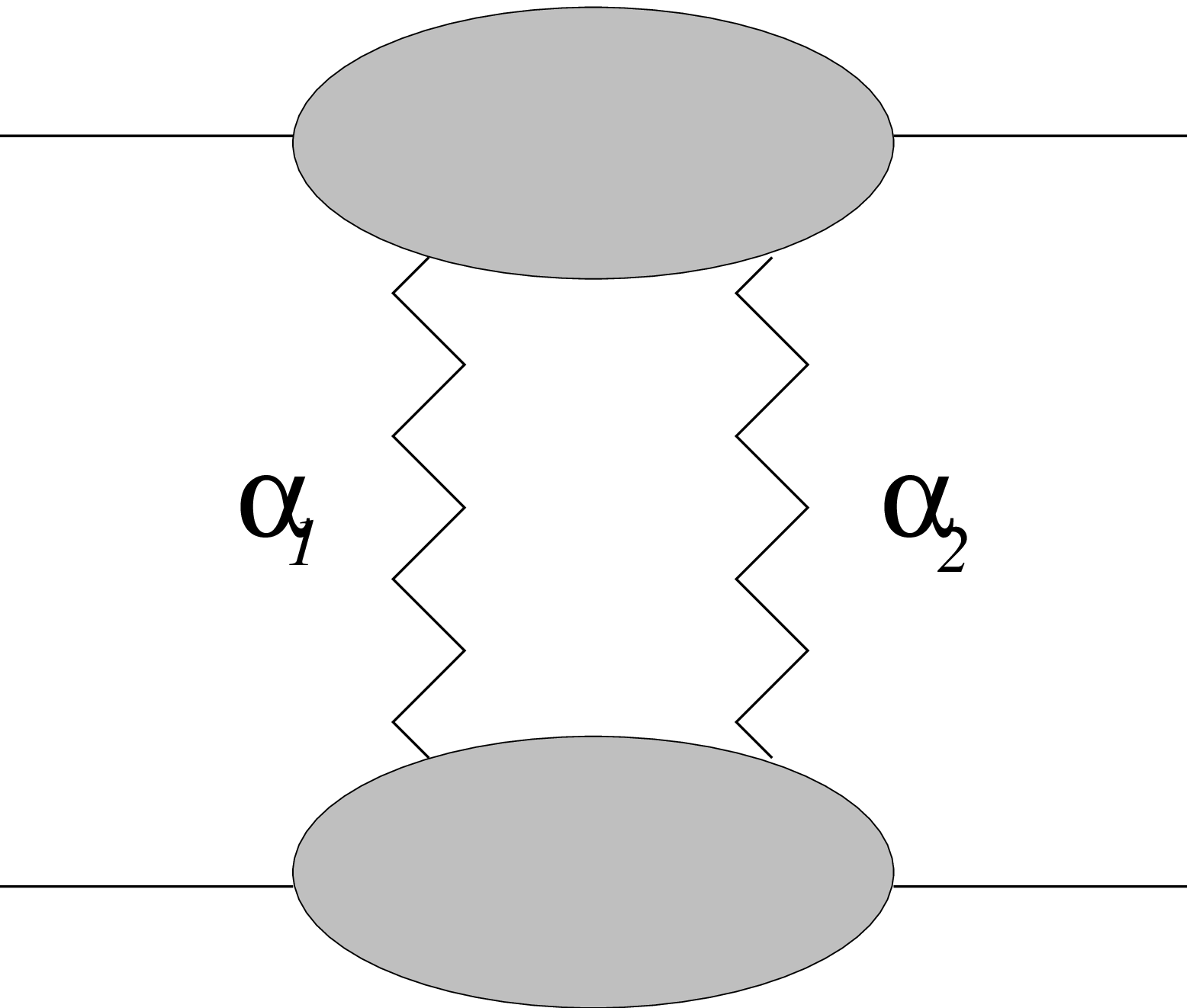}}}\hfill
\vskip -5mm
\centerline{\rfont Figure 4: double exchange}
\endinsert

A complication is that, if we can have the exchange associated with a
trajectory $\alpha (t)$, we can also have two or more such exchanges.
For example, figure 4 shows double exchange, associated with the
trajectories $\alpha _1(t)$ and $\alpha _2(t)$ (which may be the same).
This is known\ref{\collins} to give a branch point in the complex $\ell$-plane,
whose position is 
$$
\ell =\alpha _C(t)$$$$
\alpha _C(0)=\alpha _1(0)+\alpha _2(0)
\eqno(4)
$$
So the exchange of two pomerons contributes to $\sigma ^{\hbox{\sevenrm TOT}}$ a term
$s^{2\e _0}$, divided by some function of log $s$ and multiplied by some
constant which we cannot calculate, though we know that it is negative. 
The {\sl simplest assumption} is that this constant is small. Then 
the sum of the exchanges $P+PP$ will behave as an effective power
$s^{\e}$, with $\e$ just a little less than $\e_0$ and decreasing slowly
with $s$ as $s$ increases. According to the fits in figure 3,
experiment finds $\e\approx 0.08$. 

In the fits of  figure 3, the ratio of the strengths of
pomeron exchange in $\pi p$ and $pp$ or $\bar pp$ scattering is
13.6/21.7$\approx$ 2/3. This is an indication that the pomeron couples to
single valence quarks in a hadron, and is called the ``additive-quark rule''.
The {\sl simplest assumption}\defref\polk{
P V Landshoff and J C Polkinghorne, Nuclear Physics B32 (1971) 541
} is that its coupling to a quark is like that of a photon, with a
Dirac $\gamma$ matrix times a constant $\beta _0$. Then the contribution 
from pomeron exchange to the quark-quark elastic scattering amplitude is
$$
\gamma\cdot\gamma \;\beta _0^2\,s^{\alpha (t)-1}\left (-e^{-\half i\pi\alpha (t)}
\right )
\eqno(5)
$$
The last factor is the phase factor  $\xi_{\alpha (t)}$ of (2b) for the case of
charge parity $C=+1$ exchange; the inclusion of this phase 
is what makes pomeron exchange
different from photon exchange. For $pp$ or $\bar pp$ scattering, we
need to take account of the wave function of the quarks in the nucleon. Just
as for photon exchange, we do this by introducing two Dirac elastic
form factors, $F_1(t)$ and $F_2(t)$. These have been measured in $ep$ 
scattering, but there it is the photon that is exchanged, and it has $C=-1$.
The {\sl simplest assumption}, which works better than can really be
understood, is that the $C=+1$ and $C=-1$ form factors are equal. Since
pomeron exchange is isospin 0, this means that we use the sum of the
proton and neutron form factors measured in elastic electron scattering.
For the case of $F_2$, this sum is small --- at $t=0$ it is equal to the
sum of the anomalous magnetic moments of $p$ and $n$, which is small.
The presence of an $F_2$ term would correspond to nucleon helicity flip,
which has long been known to be small for pomeron excange; it is interesting 
that this can be linked to the anomalous moments\ref{\polk}. For the neutron,
the form factor $F_1$ is by definition 0 at $t=0$, and it is known to remain 
small away from $t=0$, so the form factor $F_1(t)$ that we need is just
the proton form factor $F_1(t)$ measured in elastic $ep$ scattering.
The related Sachs form factors $G_E(t)$ and $G_M(t)$ are found to be 
proportional to each other and of dipole form; the data for these correspond to
$$
F_1(t)={4m^2-2.8t\over 4M^2-t}\left ({1\over 1-t/0.7}\right )^2
\eqno(6)
$$

\midinsert
\centerline{{\epsfxsize=\hsize\epsfbox{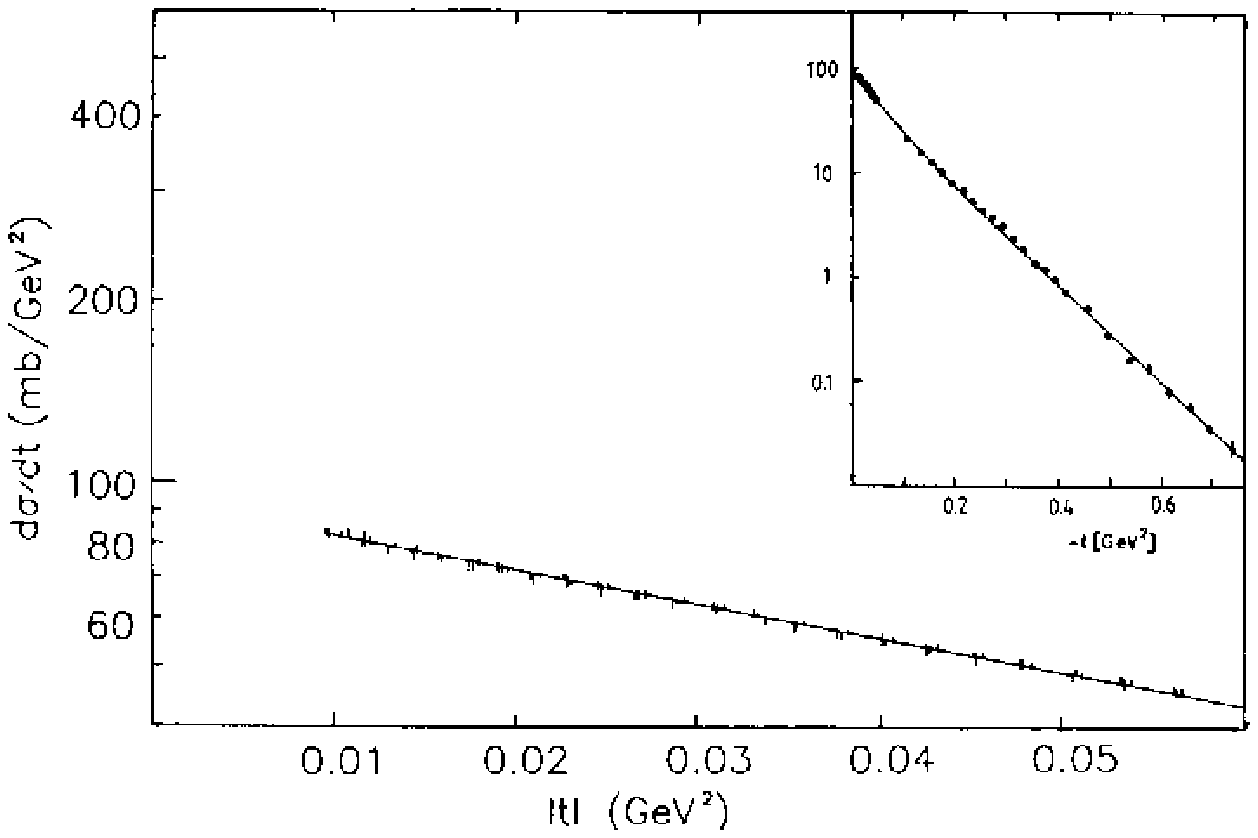}}}\hfill
\vskip -5mm
\centerline{\rfont Figure 5: $pp$ elastic scattering at $\surd s$=53 GeV}
\endinsert

Introducing the {\sl simplest assumption} that the pomeron trajectory 
$\alpha (t)$ is linear in $t$, though allowing for the possibility that it has 
a different slope $\alpha '$ from the trajectories shown in figure 2, we
find that single pomeron exchange contributes to elastic $pp$ or $\bar pp$
scattering
$$
{d\s\over dt}={[3\beta _0 F_1(t)]^4\over 4\pi} \;(\alpha 's)^{2\e _0-2\alpha '|t|}
\eqno(7)
$$
The value of $\alpha '$ may be determined by fitting this to the highly
accurate CERN ISR small-$t$ data at $\surd s$=53 GeV, shown in figure 5.
The inset in this figure shows that then the form (7) fits extremely well
to the data at larger $t$. This is a nontrivial check that the assumption
about $F_1(t)$ is surprisingly correct; as it comes into (7) raised to
the fourth power the fit is rather sensitive to it. The form (7) is
found to agree well with data at all energies\defref\dynamics{
A Donnachie and P V Landshoff, Nuclear Physics B267 (1986)  690
}, including the Tevatron data at
$\surd s=1800$ GeV. It correctly predicted that the forward peak at this
energy would be rather steeper. According to (7), if one fits to $e^{-b|t|}$
then when the energy is increased by a factor $R$ the slope $b$ decreases
by an amount $\alpha '\log R$, which is about 3.5 when the energy increases
from ISR to Tevatron values. Notice, though, that a fit with $e^{-b|t|}$
can only be local, unless one allows $b$ to vary with $t$.

Single-pomeron exchange is not the whole story. There are also nonleading
exchanges, in particular $\rho,\omega,f,a$, though these have become unimportant
when the energy is as high as 53 GeV. What cannot be ignored is the exchange
of two pomerons. While we do not know how large is the contribution from this,
we do know about its general features: see figure 6. 
It is flatter than single-pomeron exchange, and as $s$ increases it steepens
half as quickly. But at $t=0$ it rises twice as fast as single-pomeron exchange.
So, as $s$ increases the point where the two are equal moves to lower and
lower $t$. One consequence of this is that the shape of the
differential cross-section, as a function of $t$, changes with increasing 
energy. It happens that, at Tevatron energy, the two contributions combine
in such a way that a fit $e^{-b|t|}$ with $b$ independent of $t$ is quite
good\defref\esev{
E710 collaboration: N A Amos et al, Physics Letters B247 (1990) 127
}, though this is not true at either lower or higher energies.

\midinsert
\centerline{{\epsfxsize=60truemm\epsfbox{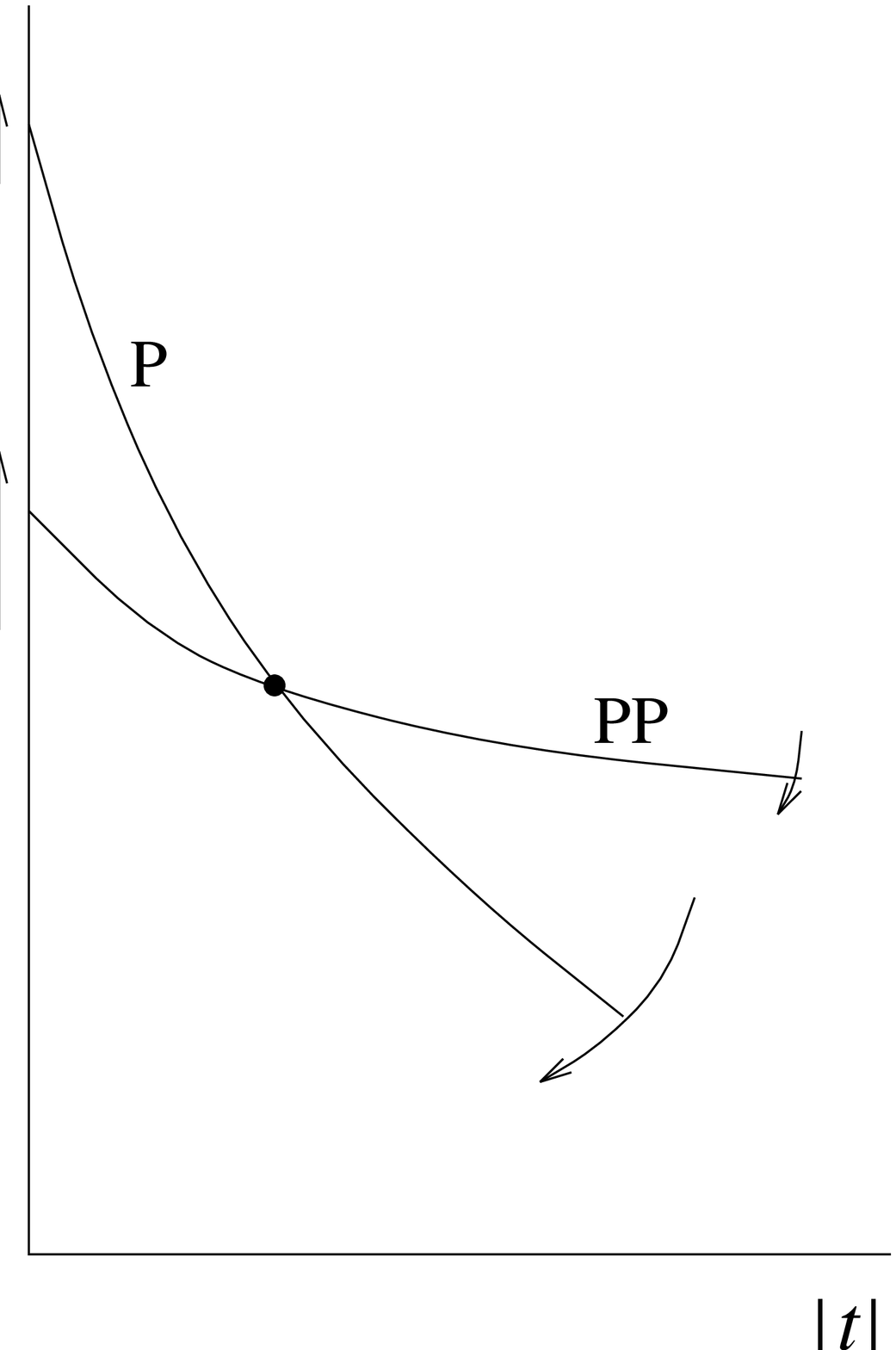}}}\hfill
\vskip -5mm
{\rfont Figure 6: contributions to ${d\s\over dt}$ from single and 
double pomeron exchange. The arrows indicate how they change as the energy
increases.}
\endinsert

Having established that, for $t<0$, the pomeron trajectory is
$$
\a (t)=\e _0+0.25 t
\eqno(8)
$$
with $\e _0$ between 0.8 and 0.9,
we may extrapolate it  to positive $t$. The {\sl simplest assumption} is that
it remains straight, and then $\a (t)=2$ at a value of $t$ just less than
4 GeV$^2$. This leads us to predict that there should be a 2$^{++}$ particle
with a mass just less than 2 GeV. Since theoretical prejudice leads to the
belief that pomeron exchange is gluon exchange, this particle would be a
glueball. It is interesting that the WA91 experiment\defref\wa{
WA91 collaboration: S Abatzis et al, Physics Letters B324 (194) 509
}
has reported a ``2$^{++}$ glueball candidate'' at just the right mass.

\midinsert
\centerline{{\epsfxsize=50truemm\epsfbox{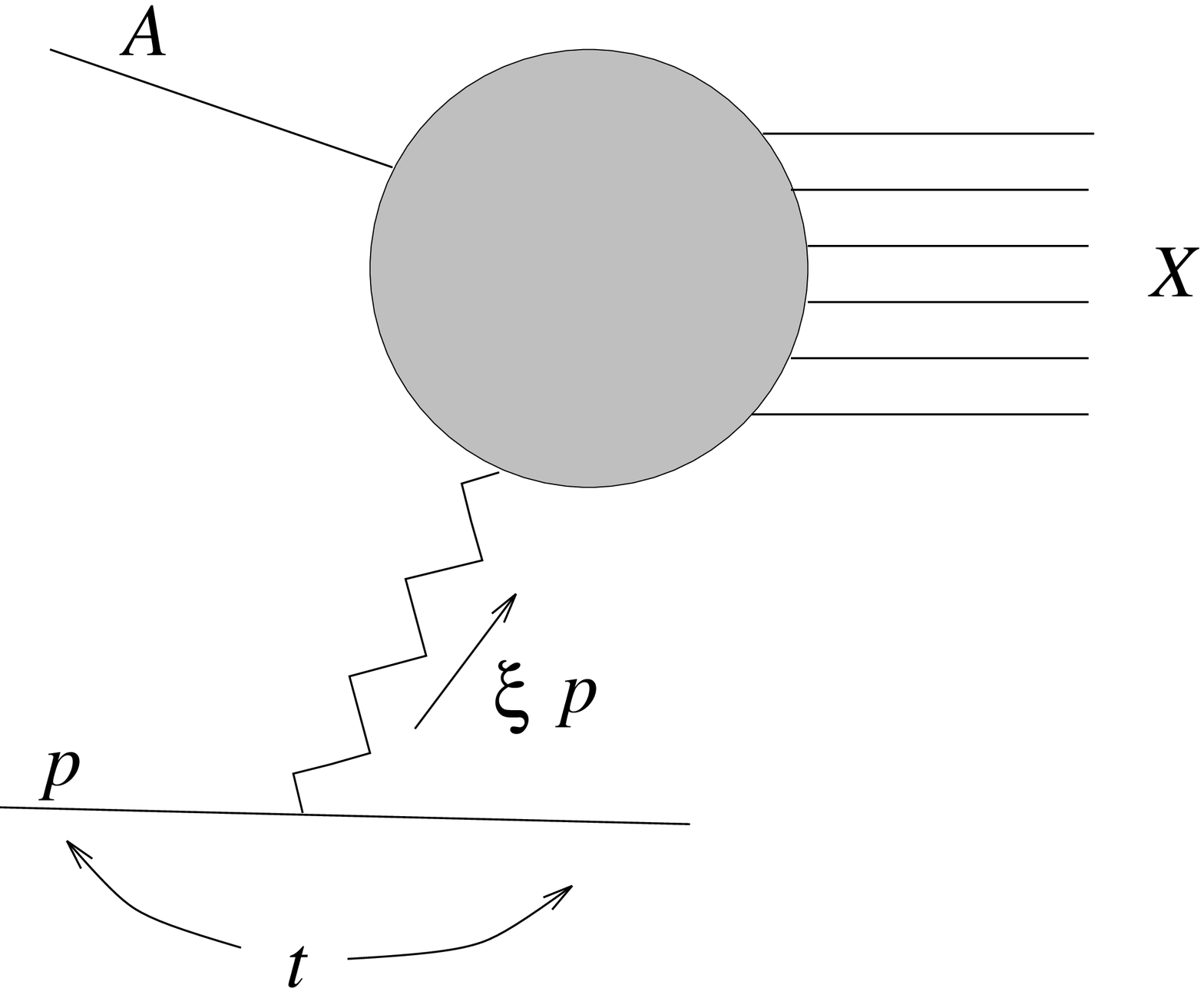}}}\hfill
\vskip -5mm
\centerline{\rfont Figure 7: diffraction dissociation}
\endinsert

\bigskip
{\bf 4. Diffraction dissociation}

Figure 7 shows diffraction dissociation: some projectile
$A$ hits a proton and breaks up into a sytem $X$ of hadrons, while the proton
survives and retains almost all its momentum. The projectile $A$ can be
any particle, for example another proton, or a $\g$ or $\g ^*$. The 
fractional momentum loss $\xi$ of the target proton should be less than a
few percent. In this case it is a matter of simple kinematics to understand
that the final state can have no other particle close in rapidity to the 
target proton, so these events are ``large-rapidity-gap'' events.
The magnitude of $\xi$ may be calculated from the invariant mass of the
system $X$ of fragments of the projectile particle: 
$\xi =M_X^2/s$. Instead of $\xi$, the notation $x_P$ is often used.

If $\xi$ is small enough, the exchanged object in figure 7 should be the
pomeron. If pomeron exchange is described by a simple pole in the complex
$\ell$-plane, it should factorise:
$$
{d^2\s ^{Ap}\over dt d\xi}=F_{P/p}(\xi ,t)\;\s ^{PA}(M_X^2,t)$$$$
F_{P/p}={9\beta _0^2[F_1(t)]^2\over 4\pi}\xi ^{1-2\a (t)}
\eqno(9)
$$
Even if there is a glueball associated with the pomeron trajectory near
$t=4$ GeV$^2$, when it is exchanged near $t=0$ the pomeron cannot be said
to be a particle. Nevertheless, the factorisation (9) makes pomeron
exchange very similar to particle exchange: the factor $\s ^{PA}(M_X^2,t)$
may be thought of as the cross-section for pomeron-$A$ scattering. When
its subenergy $M_X$ is large, it should have much the same power behaviour as
the hadron-hadron total cross-sections shown in figure 3:
$$
\s ^{PA}(M_X^2,t)\sim u(t) (M_X^2)^{0.08} + v(t) (M_X^2)^{-0.45}
\eqno(10)
$$

But there are complications: the zigzag line in figure 7 may not be the pomeron.
Simple pomeron exchange may be contaminated in two ways. If $\xi$ is not
small enough, one must add in a contribution from $\rho,\o ,f, a$ exchange,
or even $\pi$ exchange when $t$ is close to 0. That is, these exchanges
can also result in large rapidity gaps, though as they correspond to smaller
powers of $1/\xi$ than pomeron exchange, they become relatively less
important as $\xi$ decreases. If one integrates (9) down 
to some fixed $M_X^2$, the resulting cross-section for diffraction
dissociation behaves as $s^{2\e _0}$, and so unless something else intervenes
it would become larger than the total cross-section\defref\ss{
G A Schuler and T Sjostrand, Physics Letters B300 (1993) 169\h
E Gotsman, E M Levin and U Maor, Physical Review D49 (1994) 4321\h
K Goulianos, Physics Letters B358 (1995) 379
}. As $s$ increases at fixed $M_X^2$, one is probing larger and larger values
of $1/\xi$, so one expects that the same happens as in the total cross-section:
the exchange of two pomerons becomes important and moderates the rising
contribution from single exchange. But the {\sl simplest assumption} is that
this matters only at very small $\xi$.

Notice that the theory leads us to expect that 
adding these other exchanges  should give us all the nonleading powers of
$1/\xi$: there should be no other appreciable ``background''.
Note also that adding in the other exchanges will surely break factorisation.
Further, it is likely that, even though $f$ exchange, in particular, gives a 
nonleading power of $1/\xi$, it may be numerically important down to quite small
$\xi$. This certainly seems to be true for diffraction dissociation in $pp$
or $\bar pp$ collisions\defref\robroy{
D P Roy and R G Roberts, Nuclear Physics B77 (1974) 240
}. Donnachie and I parametrised\defref\diff{
A Donnachie and P V Landshoff, Nuclear Physics B244 (1984) 322
} the ISR data in the simplest
manner: we included $f$ exchange simply by multiplying the pomeron-exchange 
contribution  (9) by the factor
$$
1 +2C\xi ^a(t)\cos\half\pi a(t) +C^2\xi ^{2a(t)}$$$$
a(t)=\a _P(t) -\a _f(t) =0.64-0.68t
\eqno(11)
$$
The $\xi ^{2a(t)}$ term corresponds to the pomeron in figure 7 being replaced
with an $f$, while the $\xi ^{a(t)}$ term 
is interference between pomeron and $f$ exchange. We found that $C$ is
large, about 8, which means that at $t=0$ the factor is greater than 2
even when $\xi$ is as small as 0.02. There is no reason to suppose that
it is actually correct to use a simple factor such as (11), and the magnitude
of the effect could be substantially different for different projectiles,
such as $\g ^*$. 

The case where the projectile $A$ in figure 7 is a $\g ^*$ is what is
studied in the ``diffractive events'' at HERA. In this case, a factorising
single-pomeron exchange would  give a factorising contribution to the proton
structure function from
very-fast-proton events: 
$$
{d^2 F_2 ^{\hbox{\sevenrm DIFFRACTIVE}}\over dt d\xi}=F_{P/p}(\xi ,t)\;
F_2^{\hbox{\sevenrm POM}}(\beta,Q^2,t)
\eqno(12)
$$
where $\beta =x/\xi$. Here, $F_2^{\hbox{\sevenrm POM}}$ may be thought of
as the ``structure function of the pomeron'': it is defined if the pomeron
is a simple pole in the complex $\ell$-plane and so gives a factorising
contribution, even though it is not a particle.

According to what I have said, one has to worry about possible contamination,
particularly from $f$ exchange.  This is likely to be important if $\xi$ is
not small enough. But the value of $\xi$ below which one can forget it may well
be $\beta$-dependent. If the structure function of the $f$ is much
larger at small $\beta$ than that of the pomeron, then it might give
appreciable contamination at small $\beta$ even when $\xi$ is rather small. 

Our theoretical understanding of the pomeron structure function is so far
very rudimentary, though it did allow the prediction\defref\pomstruc{
A Donnachie and P V Landshoff, Nuclear Physics B303 (1988)  634
} that surprisingly-large
fraction of small-$x$ events at HERA would have a very fast proton in the final
state. This prediction used the simplest model, which exploits the similarity
between the pomeron and a photon, though with the important difference
that the pomeron does not couple to a conserved current. This leads
to a quark structure functions 
$$
\beta q^{\hbox{\sevenrm POM}}(\beta ) = C\beta (1-\beta )
\eqno(13)
$$
with $C\approx 0.25$ for each light quark and antiquark. A similar form
results\defref\nik{
N N Nikolaev and B G Zakharov, Z Physik C53 (1992) 331\h
M Diehl, M Diehl, Z Physik C66 (1995) 181
} from modelling pomeron exchange as two-gluon exchange. Just as for the
case of the photon structure function, one has to add in a term that is
important at small $\beta$ and behaves like $\beta ^{-\e _0}$, with
$\e _0 =0.08$, or maybe larger. This is certainly only the crudest model,
and it leaves many obvious questions. How does $q^{\hbox{\sevenrm POM}}(\beta )$
evolve\defref\gerh{
T Gehrmann and W J  Stirling, Z Physik C70 (1996) 89
} with $Q^2$? How large is the charm structure function? And what is the gluon
structure function? We have no model for the pomeron's gluon structure function,
and cannot even tell how large it should be --- as the pomeron is not
a particle, there is no momentum sum rule.

\bigskip
{\bf 5. Electroproduction of vector mesons}

So far, our theoretical understanding of the soft pomeron in terms of
QCD is only very crude. There is a consensus that pomeron exchange is
gluon exchange and that the soft pomeron is nonperturbative and so
the gluons are not perturbative. Of course, it is this which hinders any
clean calculation.

\midinsert
\centerline{{\epsfxsize=70truemm\epsfbox{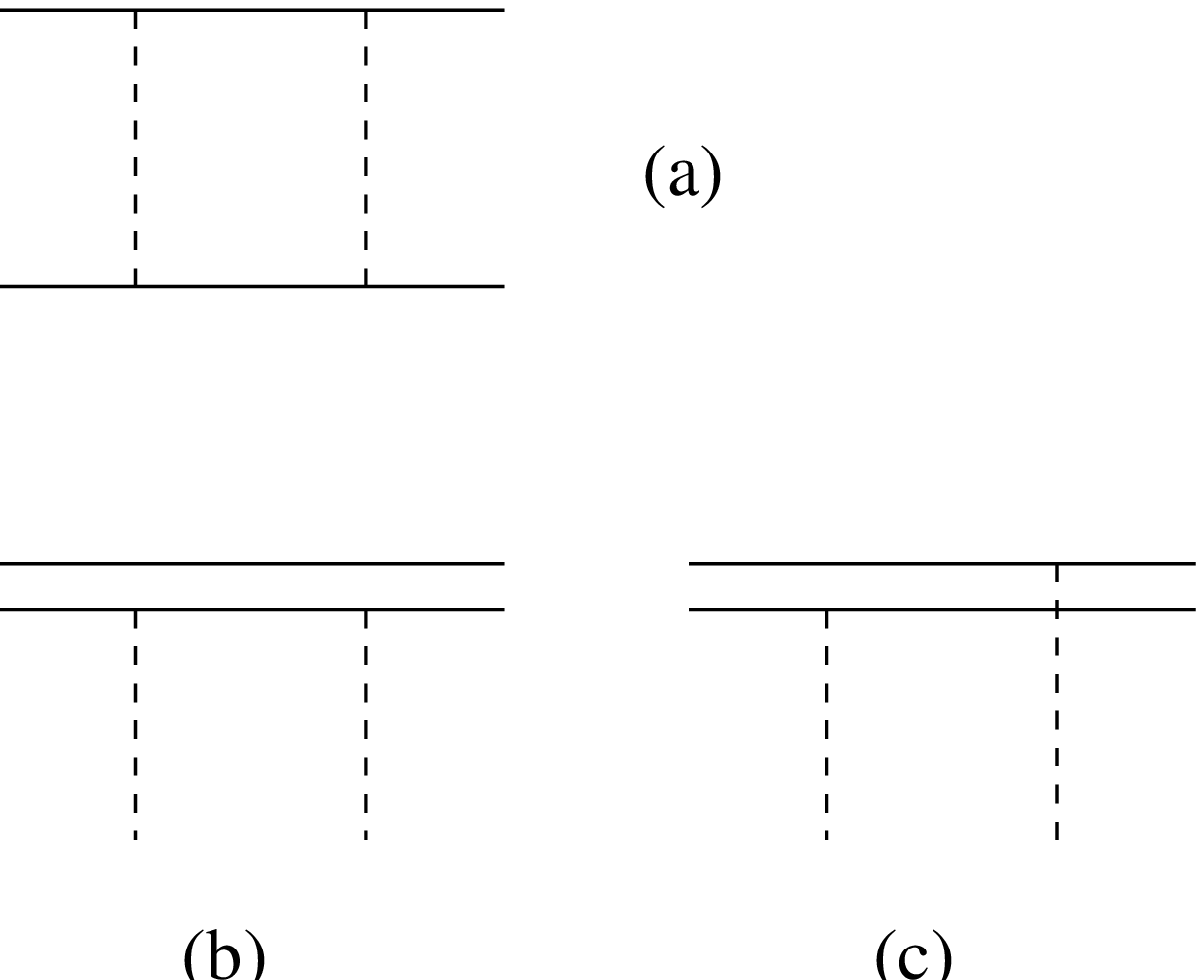}}}\hfill
\vskip -5mm
\centerline{\rfont Figure 8: (a) two-gluon between quarks; (b) and (c)
couplings of two gluons to the quuarks in a pion}
\endinsert

The gluon is confined, which means that its propagator $D(k^2)$ should
have the perturbative $k^2=0$ pole removed by nonperturbative effects.
This means that in the ratio
$$
\mu ^2={(\int _{-\infty}^0 dk^2\; 2k^2 D^2(k^2))\over
(\int _{-\infty}^0 dk^2\;  D^2(k^2))}
\eqno(12)
$$
the integrals in both the numerator and the denominator should converge
at $k^2=0$. Because confinement is a nonperturbative effect, and because
the typical nonperturbative scale is 1 GeV, we expect the mass $\mu$ defined
by (12) to be about 1 GeV. 

In order to model pomeron exchange by gluon exchange, we need at least two
gluons to reproduce the colour-singlet isoscalar nature of the pomeron.
the simplest model for pomeron exchange between quarks is thus figure 8a.
At $t=0$ this diagram is just a constant times the denominator of (12), so
the model makes sense only because of confinement. Crude as it is, the model
already has some success\defref\otto{
P V Landshoff and O Nachtmann, Z Physik C35 (1987)  405
} in explaining observed properties of the soft
pomeron: one finds from it that the two gluons couple to each quark like
a single photon-like object, with Dirac matrix $\g$.  Further, when the
two gluons couple to the quarks in a hadron, one can understand why they
prefer to couple to the same quark: in the case of a pion the coupling
of figure 8b is much larger than that of figure 8c because the pion radius
$R$ satisfies $\mu ^2 R^2 \ll 1$. Thus the additive-quark rule may be
understood.

One may refine this simple model\defref\dosch{
O Nachtmann, Annals of Physics 209 (1991) 436\h
H G Dosch and E Ferreira, Physics Letters B318 (1993) 197
} by allowing the exchange of more than two gluons, but the problem of
calculating the energy dependence $s^{\e _0}$ of pomeron exchange is still
too difficult; this factor has to be put in by hand.

\midinsert
\centerline{{{\epsfxsize=100truemm\epsfbox{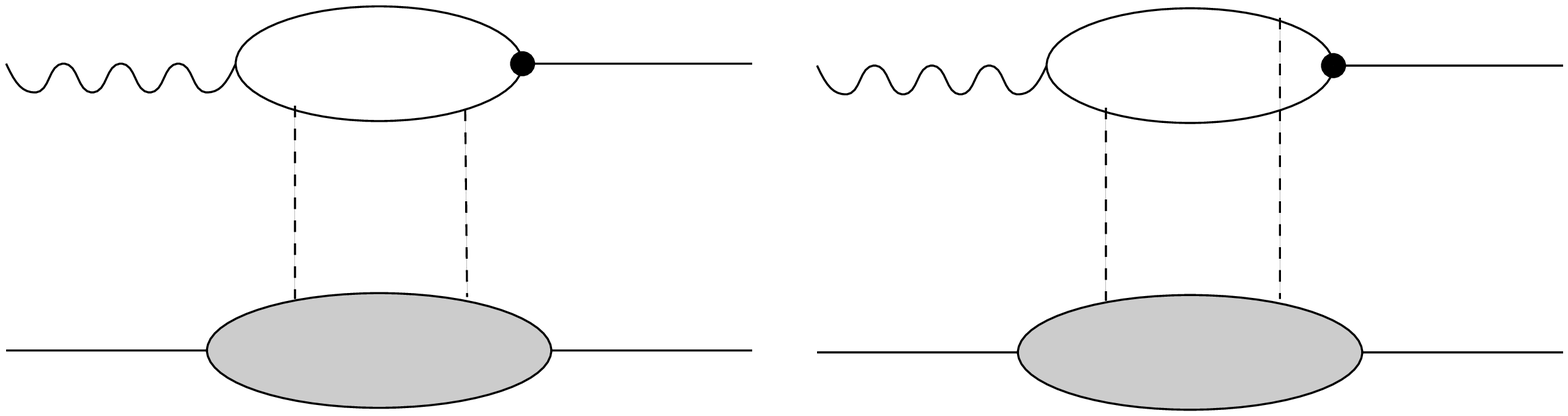}}}}\hfill\break
\centerline{\rfont{Figure 9: simple model for $\g ^*p\to\rho p$}}
\endinsert

A good test of the simple model is exclusive $\rho$ electroproduction,
$\g ^* p\to\rho p$. Apparently different approaches to this process
actually share common key features: see figure 9. At the top of each graph
is a quark loop that couples the $\gamma ^*$ to the $\rho$. There are two
different ways in which the gluons couple; the additive-quark rule
would make the first graph dominate for a real photon, but as $Q^2$
increases the second graph becomes more and more important. It tends to
cancel the first graph: this is called `colour transparency''. 
As for the bottom bubble in the graphs, one approach is to 
pretend\defref\brodsky{  
S Brodsky et al, Physical Review  D50 (1994) 3134
} that it is the gluon structure function of the proton, though in fact it
cannot exactly be that and indeed the assumption could be altogether 
wrong\defref\vector{
A Donnachie and P V Landshoff, Physics Letters B348 (1995) 213
}. Using the gluon structure function leads to a rapid rise with increasing
energy $W$. 

\midinsert
\vskip -41truemm
\centerline{{{\epsfxsize=100truemm\epsfbox{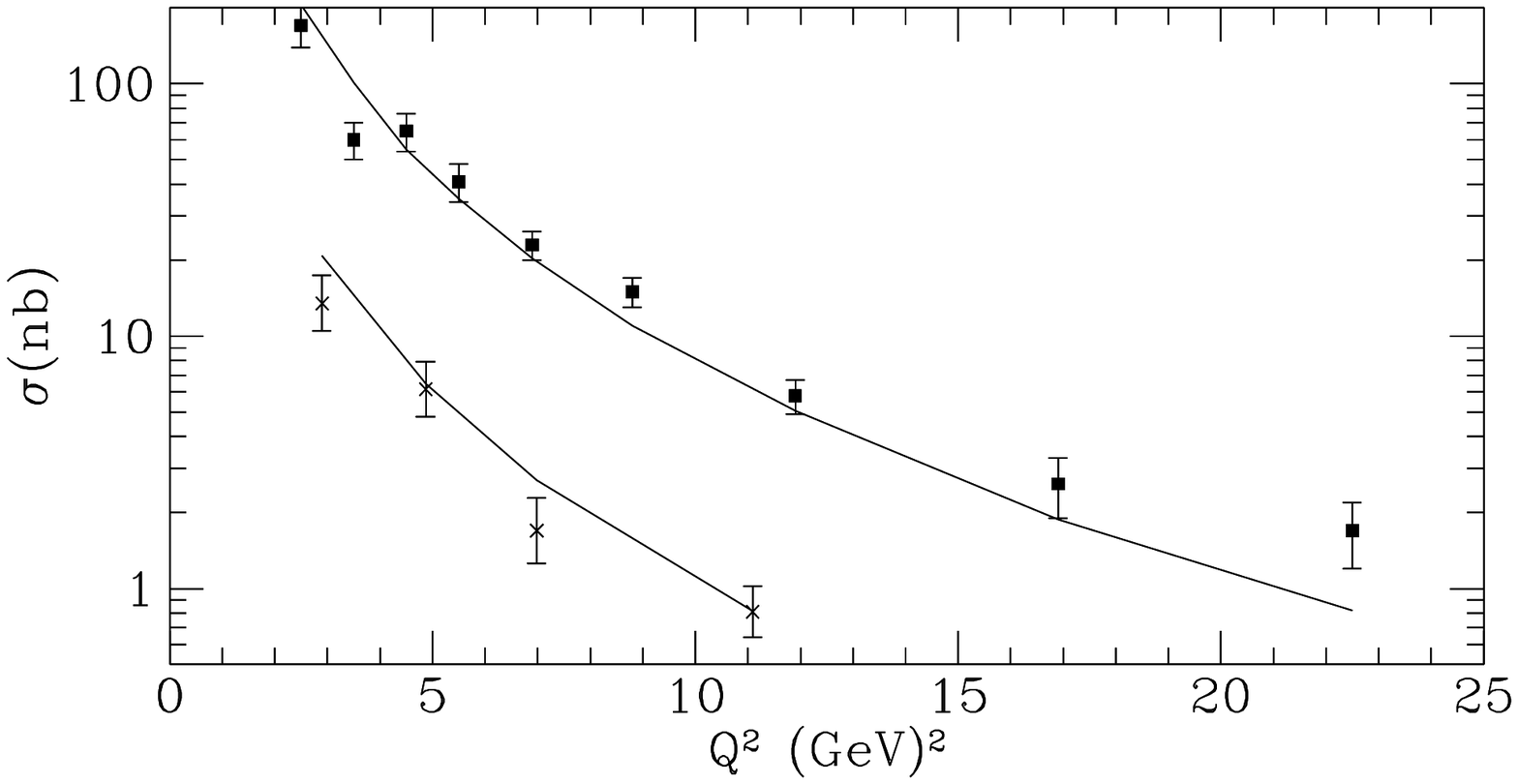}}}}\hfill\break
\vskip -11truemm
{\rfont{Figure 10: NMC data\defref\nmc{
NMC collaboration: P Amaudruz et al, Z Physik C54 (1992) 239
} for $\g ^* p\to\rho p$ and $\g ^* p\to\phi p$, with calculated curves
from reference {\vector}}}
\endinsert

A simpler model\defref\rrho{
A Donnachie and P V Landshoff, Nuclear Physics B311 (1989) 509\h
J R Cudell, Nuclear Physics B336 (1990) 1
} is to replace the bottom bubble with the same simple coupling to a
quark as in figure 8. Then the $W$ dependence has to be put in by hand. 
This model is surprisingly successful in its agreement with low-energy
data: see figure 10, which includes also $\g ^*p\to\phi p$. The calculated
curves shown in figure 10 also make a simple assumption about the $\rho$ vertex:
that it is strongly peaked such that the two quarks at the vertex prefer to
share equally the momentum of the $\rho$.  A test of this is the 
consequence for the $\rho$ polarisation (which should be equal to that
of the $\g ^*$. For $Q^2\gg m^2_{\rho}$
the longitudinal amplitude is proportional to $f_{\rho}/Q^3$ and the 
transverse amplitude to $f_{\rho}m_{\rho}/Q^4$.  The ratio of the amplitudes
is predicted to be about 2 for $Q^2=5$ GeV$^2$, rising to 8 at $Q^2=20$,
which seems to be in agreement with the low-energy data.
Because of the extra factor $m/Q$ in the transverse amplitude, for
heavier vector mesons we expect to need rather larger $Q^2$ before the
longitudinal production dominates. Thus for $\g ^*p\to J/\psi\, p$, the 
simple model predicts that we have to go to $Q^2$=100 before the longitudinal 
amplitude is twice as large as the transverse. Nevertheless the transverse 
amplitude is big if the coupling of the (nonperturbative) gluons to the
quarks is flavour-blind: $J/\psi$ production overtakes $\rho$ production
at around $Q^2$=10. How much these predictions depend on the explicit
assumptions about the vertex is not understood.

\bigskip
{\bf 6. Conclusions}

The soft pomeron successfully correlates a variety of $Q^2=0$ data. Its 
properties are probably simple -- it seems to couple to single quarks
in a factorising manner, indicating that it is associated with a simple pole
in the complex $\ell$-plane.

Nevertheless, there are some big surprises in the HERA data. The cross-section
for quasi-elastic $J/\psi$ photoproduction, $\g p\to J/\psi\, p$, rises
more rapidly with energy than soft-pomeron exchange would have predicted, and
the proton structure function $F_2$ rises spectacularly rapidly as $x$
becomes very small. An immediate explanation that comes to mind is that
one is seeing the effects of the perturbative BFKL pomeron\defref\bfkl{
E A Kuraev, L N Lipatov and V S Fadin: Sov.Phys.JETP44 (1976) 443
}, but this is unlikely to be correct\defref\us{
J R Cudell, A Donnachie and P V Landshoff, hep-ph/9602284
}. There are several other 
candidate explanations\defref\capella{
A Capella et al, Physics Letters B337 (1994) 358\h
M Gl\"uck, E Reya and A Vogt, Z Physik C67 (1995) 433\h
R D Ball and S Forte, Physics Letters B351 (1995) 313
}, but no general agreement about what is the right one.
As was said by Uri Maor at the recent meeting in Eilat:

``One of the reasons it is a beautiful subject is that there are lots of 
things we don't understand''.

\bigskip\bigskip
{\it This research is supported in part by the EU Programme ``Human Capital
and Mobility", Network ``Physics at High Energy Colliders'', contract
CHRX-CT93-0357 (DG 12 COMA), and by PPARC.}
\bigskip
\bigskip\medskip\immediate\closeout\rfile\writestoppt
\baselineskip=10pt{{\bf References}}\bigskip{\frenchspacing%
\parindent=3truepc\escapechar=` \input refs.tmp\bigskip}\nonfrenchspacing
\bye